\documentclass[twocolumn]{revtex4-1}   	
\usepackage{geometry}                		
\usepackage{amsmath, amssymb, graphics, setspace}
\newcommand{\mathsym}[1]{{}}
\newcommand{\unicode}[1]{{}}

\usepackage{graphicx}				
\usepackage{epsfig}				         
\newcommand{\beq}{\begin{eqnarray}}
\newcommand{\eeq}{\end{eqnarray}}
\begin{document}
\title{Regions  of Applicability of Approximate Formulations of Neutrino Oscillations in Matter }
\author{
Mikkel B. Johnson \\
Los Alamos National Laboratory, Los Alamos, NM 87545 \\
Ernest M. Henley\\
Department of Physics, University of Washington, Seattle, WA 98195\\ 
Leonard S. Kisslinger\\
Department of Physics, Carnegie-Mellon University, Pittsburgh, PA 15213\\
}
\begin{abstract}

We examine the reliability of the theoretical methods presently used for analysis and prediction of neutrino oscillation phenomena. Of particular interest are the limitations imposed by branch points when expansions in one of the small parameters of the Standard Neutrino Model (SNM) are made to obtain tractable results. Our evaluation compares the approximate oscillation probabilities for flavor-changing transitions in the $(\nu_e,\nu_\mu)$ sector to exact results obtained from of a recently- developed exact analytical representation of neutrino oscillations in matter within the SNM. From our numerical comparisons, we are able to identify regions where the existing approaches can be improved to take full advantage of the higher quality data expected at future neutrino facilities.

\end{abstract}
\maketitle
\noindent
PACS Indices:  11.3.Er,14.60.Lm,13.15.+g 
\vspace{1mm}
\noindent

Keywords:

\vspace{1mm}
\noindent

\section{Introduction}

In this paper, we explore the theoretical errors of simple expressions for  oscillation probabilities found from approximate formulations for the purpose of analysis and prediction of experiment at current and proposed future neutrino facilities. Our assessment  includes the interaction of neutrinos with electrons shown to be essential by Wolfenstein in 1978~\cite{wolf} and later identified by Mikheyev and Smirnov~\cite{SMAS} as a likely explanation of the deficit of solar neutrinos discovered experimentally by Davis~\cite{Davis,Clev}.  Subsequent to the work of Wolfenstein, the effect of matter on neutrino oscillations has been  commonly explored using exact computer simulations.  Computer simulations of neutrino oscillations in matter  are found in Refs.~\cite{MS} and references therein.  Such numerical studies are preferred when density profiles with a spatial dependence are essential.  

Our assessment of expressions for the oscillation probability in the flavor-changing $(\nu_e,\nu_\mu)$ sector is made by comparing the approximate expressions  to  corresponding  exact analytical expressions for the oscillation probability of  3 coupled Dirac neutrinos.   The expressions assessed include not only neutrino oscillation probabilities found in  Refs.~\cite{f,ahlo,JO} but also neutrino oscillation probabilities from yet another approximate formulation we propose based on  expressions for the oscillation probability found in our recently published paper~\cite{jhk0} evaluated with $\xi$-expanded eigenvalues.

The exact analytical expressions for the oscillation probability used for the comparison are obtained from our recently-developed Hamiltonian formulation~\cite{jhk0}.  They  are algebraic  functions of the parameters of the the Standard Neutrino Model~\cite{ISS}  (SNM), the neutrino eigenvalues, and  include matter effects.   The expressions were  obtained by using the Lagrange interpolation formula~\cite{barg} to exponentiate the neutrino Hamiltonian and have been used to verify our own exploratory studies~\cite{hjk1,khj2,hjk1a,khj1E} that used the expressions appearing  in Refs.~\cite{f,ahlo,JO}.

In Sect.~\ref{pxPOP}, we review the simple analytical results presently available~\cite{f,ahlo,JO}, and in  Sect.~\ref{ExpComp}, we discuss the assessment itself.  Then, in Sect.~\ref{NCall}, we assess the oscillation probabilities found in Refs.~\cite{f,ahlo,JO}. This assessment, which appears  in Sect.~\ref{Xcompare},  confirms  the finding of Ref.~\cite{jhk0} that the  accuracy of the approximate analytic results of Refs.~\cite{f,ahlo,JO} is limited by the presence of branch points in the analytic structure of the eigenvalues of neutrinos propagating in matter.     

The analysis in Sect.~\ref{Xcompare} also suggests that the the simple expressions  for neutrino oscillation probability we find  using  our Hamiltonian formulation~\cite{jhk0} are accurate to a few percent in all  regions.   It also  identifies regions where the approximate analytical results are reliable by making numerical comparisons to exact results~\cite{jhk0}. The approximate expressions we propose  are  conservatively estimated there to be accurate to a few percent  within essentially all regions of interest.  Approximate expressions of  such accuracy would, of course,  obviate the need for exact  computer simulations under many circumstances.

Combining the results of Sect.~\ref{NCall}  with numerical results presented in Ref.~\cite{jhk0}, we identify the regions where the approximate  oscillation probabilities found from our proposed formulation are more accurate than those of Refs.~\cite{f,ahlo,JO}. In a future publication~\cite{jk1}, we show how to construct such approximate oscillation probabilities in practice guided by the assessment made here.

\section { Neutrino Dynamics }
\label{ND}

The dynamics of the three known neutrinos and their corresponding anti-neutrinos in matter is determined by the time-dependent Schroedinger equation,
\begin{eqnarray}
\label{tdse}
i\frac{d}{dt}|\nu(t)> &=& H_\nu |\nu(t)> ~,
\end{eqnarray}
where the neutrino Hamiltonian $H_\nu$, 
\beq
\label{fulh}
 H_\nu=H_{0v}+H_{1} ~,
\eeq
consists of a piece $H_{0v}$ describing neutrinos in the vacuum and a piece $H_{1}$ describing their interaction with matter.

The solutions of Eq.~(\ref{tdse})  may be expressed in terms of stationary-state solutions of the eigenvalue (EV) equation
\beq
\label{eveq}
H_\nu|\nu_{mi}> &=& E_i |\nu_{mi}> ~,
\eeq
where the label ``$m$" indicates neutrino mass eigenstates, as distinguished from their flavor states sometimes denoted the label ``$f$".  
Neutrinos are produced and detected in states of good flavor.

In operator form, the dynamics of neutrinos may be expressed in terms of the time-evolution operator $S (t',t)$, which describes completely the evolution of states from time $t$ to $t'$ and also satisfies the time-dependent Schroedinger equation.

We will examine neutrinos propagating in a uniform medium for interactions constant not only in space but also time. Because the Hamiltonian is then translationally invariant, attention may be restricted to states, both in the vacuum and in matter, characterized by momentum $\vec p$ and therefore having the overall $r$-dependence $e^{i\vec p\cdot\vec r}$. In this case, expressions may be simplified by suppressing the overall plane wave, a convention we adopt. 

For time-independent interactions, $S(t',t)$,
\beq
\label{smtx}
S(t',t) &=& e^{-iH_\nu (t' - t)} ~,
\eeq
depends on time  only through the time {\it difference} $t'-t$. Then, written in terms of the stationary state solutions $|\nu_{mi}>$ of Eq.~(\ref{tdse}), 
\beq
\label{tunif}
S(t',t) &=& \sum_i |\nu_{mi}> e^{-iE_i (t' - t ) }<\nu_{mi}| ~.
\eeq

We assume here that that neutrinos and anti-neutrinos represented by $|\nu^0_{mi}>$ and $|{\bar \nu}^0_{mi}>$, respectively, are the structureless elementary Dirac fields of the the Standard Neutrino Model~\cite{ISS}.  For this reason the theory is invariant under CPT, so the mass of an anti-neutrino in the vacuum is the same as that for its corresponding neutrino.

The case of main interest for many situations  is  
the ultra-relativistic limit, $\vec |p| >> m^2$ (we take the speed of light $c=1$).  For ultra-relativistic neutrinos in the laboratory frame, the energy of a neutrino in the vacuum becomes
\beq 
\label{Evac1}
E^0_i &\approx&  |\vec p| + \frac{m_i^2}{2E} ~,
\eeq
where $m_i$ is its mass the vacuum.  Similarly, $E_i$ appearing in Eq.~(\ref{eveq}) may be written
\beq 
E_i &\approx&  |\vec p| + \frac{M_i^2}{2E} ~,
\eeq
where $M_i$ is its mass in the medium.    
Thus, in this limit and in dimensionless variables, 
\beq
{\hat {\bar E}}_i &\to& \frac{M_i^2 - m^{2}_1}{m^{2}_3 - m^{2}_1}
\eeq
and 
\beq 
\label{h0st2}
{\hat {\bar H}}_{0v} &\to & \left( \begin{array}{ccc} 0 & 0 & 0 \\ 0 & \alpha
& 0 \\ 0 & 0 & 1 \end{array} \right ) 
\eeq
with
\beq
\label{aldef}
\alpha \equiv \frac{m_2^2-m_1^2}{m_3^2-m_1^2} ~.
\eeq

In this limit, the distance $L$ from the source to the detector corresponding to $S(t',t)$ in Eq.~(\ref{smtx}) is
\beq
\label{Ldef}
L &=& t'-t ~.
\eeq
The time-evolution operator, Eq.~(\ref{smtx}), expressed in dimensionless variables is then,
\beq
\label{smtx1}
S(L) &=& e^{-i  H_\nu (t' - t)} \nonumber \\
&=& e^{2 i {\hat {\bar E}}^0_1 \Delta_L}  e^{-2 i  {\hat {\bar H}}_\nu \Delta_L} ~,
\eeq
where ${\hat {\bar H}}_\nu $ is the full neutrino Hamiltonian expressed in dimensionless variables, and where 
\beq
\label{Deldeff}
\Delta_L &\equiv& \frac{L(m^2_3-m^2_1)}{4E} ~.
\eeq

\section{The Neutrino Interaction and the Standard Neutrino Model}

Flavor and mss states  are related by the neutrino analog of the familiar CKM unitary matrix $U$,
\beq
\nu_f &=& U \nu_m ~.
\eeq  
The matrix $U$ is often parametrized in terms of three mixing angles $(\theta_{12},\theta_{13},\theta_{23})$ and a phase $\delta_{cp}$ characterizing $CP$ violation, taking the form,
\beq
U &\equiv& \left( \begin{array}{ccc} c_{12} c_{13} & s_{12} c_{13} & s_{13} e^{-i\delta_{cp}} \\ U_{21} & U_{22} & s_{23} c_{13} \\ U_{31}  & U_{32} & c_{23} c_{13}  \end{array} \right ) ~,
\eeq 
where,
\beq
\label{Udef}
U_{21}&=& -s_{12} c_{23} - c_{12} s_{23} s_{13} e^{i\delta_{cp}} \nonumber \\
U_{22}&=& c_{12} c_{23} - s_{12} s_{23} s_{13} e^{i\delta_{cp}} \nonumber \\
U_{31}&=& s_{12} s_{23} - c_{12} c_{23} s_{13} e^{i\delta_{cp}} \nonumber \\
U_{32}&=& - c_{12} s_{23} - s_{12} c_{23} s_{13} e^{i\delta_{cp}}  ~,
\eeq
and where the index  $i=1$ in $U_{ij}$ corresponds to the electron ($e$) neutrino, $i=2$ to the muon ($\mu$) neutrino, and $i=3$ to the tau ($\tau$) neutrino.
The standard abbreviations, $s_{12}\equiv\sin{\theta_{12}}$, $c_{12}\equiv\cos{\theta_{12}}$, {\it etc} have been used.

The perturbing Hamiltonian  $H_1$ is determined by the interaction between electron neutrino flavor states and the electrons of the medium.  Expressed as matrix, 
\beq
\label{vdef}
H_{1} &=& U^{-1} \left( \begin{array}{ccc} V & 0 & 0 \\ 0 & 0  & 0 \\ 0 & 0 & 0   \end{array} \right ) U ~,
\eeq
with $V = \pm\sqrt{2} G_F n_e $ and $n_e$ the electron number density in matter. 

For electrically neutral matter consisting of protons, neutrons, and electrons, the electron density $n_e$ is the same as the proton density $n_p$, 
\beq
n_e &=& n_p \nonumber \\
&=& R A ~,
\eeq
where $A=N+Z$ is the average total nucleon number density and $R=Z/A$ is the average proton-nucleon ratio. In the earth's mantle, the dominant constituents of matter are the light elements so $R \approx 1/2$; in the surface of a neutron star $R<<1$.  Matrix elements of $H_1$ are thus
\beq
\label{mxh1h0}
<M(k)|H_{1}|M(k')> &=& U^*_{1k} V U_{1k'} ~.
\eeq

Using the well-known expression for $V$, we find the corresponding the (dimensionless) interaction strength ${\hat A}$ of neutrinos and anti-neutrinos with matter to be,
\beq
\label{ahatrho}
\hat A &=& \pm 6.50  ~10^{-2} R  ~E[{\rm GeV}] \rho[{\rm gm/cm}^3] ~,
\eeq
with E[GeV] being the neutrino beam energy $E$ (in GeV) and $\rho$[gm/cm$^3$] the average total density (in ${\rm gm/cm}^3$) of matter  through which the neutrino beam passes on its way to the detector. For our experiments close to the earth's surface, the appropriate density is mean density of the earth's mantle,
\beq
\rho[{\rm gm/cm}^3] &=& \rho_0 \nonumber \\
&\approx& 3 ~.
\eeq

We adopt the SNM~\cite{ISS}, given next, to complete our description of the  neutrino Hamiltonian.  Most of the parameters of the SNM are consistent with global fits to neutrino oscillation data with relatively good precision~\cite{hjk1,khj2}.  These include the neutrino mass differences,
\beq
m^2_2-m^2_1 &\equiv& \delta m_{21}^2 \nonumber \\
&=& 7.6\times 10^{-5} ~{\rm eV}^2
\eeq
and
\beq
m^2_3-m^2_1 &\equiv& \delta m_{31}^2 \nonumber \\
&=& 2.4\times 10^{-3} ~{\rm eV}^2 ~, 
\eeq
which corresponds to
\beq
\alpha &\equiv& \frac{\delta m_{21}^2}{\delta m_{31}^2} \nonumber \\
&=& 3.17\times 10^{-2} ~.
\eeq

The mixing angles $\theta$ are also determined from experiment.  In the SNM, the
value of $\theta_{23}$,
\beq
\theta_{23}&=& \pi/4 ~,
\eeq
is the best-fit value from Ref.~\cite{Dav}, and $\theta_{12}$,
\beq
\theta_{12}&=& \pi/5.4 ~,
\eeq
is consistent with the recent analysis of Ref.~\cite{Gon}. 
The mixing angle $\theta_{13}$ is known to be small ($\theta_{13}<0.18$ at the 95\% confidence level) but until recently its precise value is uncertain.  Results  from the Daya Bay project~\cite{DB} have measured its value quite acurartely,  $\sin{\theta_{13}} \approx 0.15$, which we adopt to determine our value for $\theta_{13}$,
\beq
\theta_{13}&=& 0.151  ~.
\eeq
This fixes  $R_p \equiv \sin^2\theta_{13}/\alpha \approx 0.711$. The CP violating phase is not known at all and will one of the major interests at future neutrino facilities.

Using  the value of  $ \delta m_{21}^2$ from  the SNM, $\Delta_{L}$, defined in Eq.~(\ref{Deldeff}) becomes,  in the high-energy limit,
\beq
\label{Deldefn}
\Delta_L &\approx&  3.05 \times 10^{-3} \frac { L[{\rm Km}] }{E[{\rm GeV}]} ~.
\eeq 
Here $ L[{\rm Km}] $ is  the baseline and $E[{\rm GeV}]$ is the neutrino beam energy

\section{Exact Analytic Expressions for $\mathcal{P}(\nu_i  \rightarrow \nu_f)$ }

Our  Hamiltonian formulation formulation of neutrino oscillations~\cite{jhk0} leads to exact, closed-form, analytical expressions neutrino oscillation probability expressed as the sum of four partial oscillation probabilities,
\beq
\label{Pmmemu0}
\mathcal{P}(\nu_a  &\rightarrow&  \nu_b) =  \delta(a,b)  + P_{\sin\delta}^{ab} + P_0^{ab}  + P_{\cos\delta}^{ab} \nonumber \\
&+&  P_{\cos^2\delta}^{ab} ~,
\eeq
representing the  dependence of the oscillation probability on the CP violating phase $\delta_{cp}$.  The partial oscillation probabilities are given  in Appendix~\ref{SLTH}, where they are expressed as explicit  functions of the neutrino medium modified neutrino  eigenvalues ${\hat {\bar E}}_\ell$ and a set of coefficients $w_{i;\ell}^{ab}$.

Equations determining the exact eigenvalues, and various approximations to them, appear in Appendix~\ref{diagH}.  Finally, in Appendix~\ref{wcoef}, explicit algebraic  expressions for the coefficients $ w_{i;\ell}^{ab} $ for all transitions $\nu_a \leftrightarrow \nu_b$ are given in terms of  the parameters of the SNM.

\section{  Perturbative Expansions of $\mathcal{P}(\nu_i  \rightarrow \nu_f)$  }
\label{pxPOP}

The fact that $\alpha$ and $\sin\theta_{13}$ are naturally small in the SNM commonly motivates approximation schemes~\cite{f,ahlo,JO,Cer} based on first-order Taylor series expansions in one of these  small parameters, $\xi_i'$ (where $\xi_i'$ stands for $\alpha$ or $\sin^2{\theta_{13}})$.  For example, in Refs.~\cite{ahlo,JO} the oscillation probability is expanded in $\sin^2\theta_{13}$. Reference~\cite{f} makes use of an  expansion in the small parameter $\alpha$. In both publications, the only transitions considered are those  in the ($e, \mu$) sector.  

Although the expansions have the advantage of simplifying the theory, their use comes  at a price:  neither expansion gives an accurate representations for all regions of the interaction strength ${\hat A}$, including some regions of critical importance.  The origin of this inaccuracy is the presence of branch points in the neutrino eigenvalues~\cite{jhk0}. We refer to these  truncated $\xi_i$-expanded oscillation probabilities of Refs.~\cite{f,ahlo,JO} as ``full" results. 
In addition to his full result, Freund also introduces an ${\it ad~hoc}$ ``patched" result to repair spuriosity of  different origin.

In this section, we review  the results for the oscillation probability in the ($e, \mu$) sector obtained in Refs.~\cite{f,ahlo,JO,Cer}.  In later sections, we then compare their oscillation probabilities numerically to the corresponding exact expressions  obtained from our Hamiltonian formulation  evaluated with both the exact neutrino eigenvalues and with the $\xi_i$-expanded eigenvalues.

\subsection{ Full Result of Freund  }

In his work, Freund gives~\cite{f}  two versions of the  oscillation probabilities based on this $\alpha$-expansion, a full version and a patched version.
As in our own work, the oscillation probabilities are expressed as a functional of the flavor mixing angles and  the CP-violating phase that define the neutrino CKM mixing matrix, the neutrino mass differences squared, and the strength of the neutrino interaction with matter. 
Consequently, Freund makes simplifications~\cite{f} in two stages.  The first stage is to expand the eigenvalues, and the second stage is to use these results in an  expansion of the oscillation probability.  

Freund's $\alpha$-expanded eigenvalues, given in Eqs.~(18,19) of Ref.~\cite{f} are identical to those of Eq.~(\ref{evdiff1Y}) in the Appendix.
Numerical comparison to the exact result confirms that ${\hat {\bar E}}^{\alpha}_1$ and ${\hat {\bar E}}^{\alpha}_{2} $ are poor representations of the corresponding exact results in the vicinity of the branch point at ${\hat A}  = 0$.   

Freund's full $\alpha$-expanded partial oscillation probabilities in the $\nu_e \to \nu_\mu$ sector are,

\begin{widetext}

\beq
\label{expopal}
P^{e\mu}_0 ( \Delta_L,{\hat A} ) &=& \sin^2{\theta_{23}} \frac{\sin^2{2 \theta_{13} }}{{\hat C}_\alpha^2}  \sin^2{\Delta_L {\hat C}_\alpha } \nonumber \\
P^{e\mu}_1(\Delta_L,{\hat A}) &=& -\alpha \frac{1-{\hat A} \cos{2\theta_{13}} }{{\hat C}_\alpha^3}  \sin^2{\theta_{12} } \sin^2{2 \theta_{13} }  \sin^2{\theta_{23}} \Delta_L \nonumber \\
&\times& \sin{2{\hat C}_\alpha \Delta_L} + \alpha \frac{ 2 {\hat A}(-{\hat A}+ \cos{2 \theta_{13}} ) } { {\hat C}_\alpha^4 } \sin^2{ \theta_{12}} \sin^2{2 \theta_{13}}   \sin^2{ \theta_{23}}  \sin^2{\Delta_L{\hat C}_\alpha }   \nonumber \\P^{e\mu}_{\sin\delta}(\Delta_L,{\hat A}) &=& \frac{\alpha}{2} \sin{\delta_{cp} } \frac{ \cos{\theta_{13}} \sin{2 \theta_{12}} \sin{2 \theta_{13}} \sin{2 \theta_{23}} } { {\hat A} {\hat C}_\alpha \cos^2{ \theta_{13}  } }  \sin{{\hat C}_\alpha \Delta_L} \nonumber \\
&\times& [ \cos{{\hat C}_\alpha \Delta_L} - \cos{(1+{\hat A}) \Delta_L } ] \nonumber \\
P^{e\mu}_{\cos\delta}(\Delta_L,{\hat A}) &=& \frac{\alpha}{2} \cos{\delta_{cp} } \frac{ \cos{\theta_{13}} \sin{2 \theta_{12}} \sin{2 \theta_{13}} \sin{2 \theta_{23}} } { {\hat A} {\hat C}_\alpha \cos^2{ \theta_{13} } } \sin{\Delta_L{\hat C}_\alpha } \nonumber \\
&\times& [\sin{(1+{\hat A}) \Delta_L } \mp \sin{{\hat C}_\alpha \Delta_L}] 
\nonumber \\
P^{e\mu}_2( \Delta_L,{\hat A}) &=& \alpha \frac{\mp 1 + {\hat C}_\alpha \pm{\hat A} \cos{2 \theta_{13} } }{2 {\hat C}_\alpha^2 {\hat A} \cos^2{\theta_{13} } }  \cos{ \theta_{13}} \sin{ 2 \theta_{12}} \sin{ 2 \theta_{13}} \nonumber \\
&\times& \sin{2\theta_{23}} \sin^2{ {\hat C}_\alpha \Delta_L } \nonumber \\
P^{e\mu}_3(\Delta_L,{\hat A}) &=& \alpha^2 \frac{ 2 {\hat C}_\alpha \cos^2{\theta_{23}} \sin^2{2 \theta_{12}} }{{\hat A}^2 \cos^2{\theta_{13} } (\mp {\hat A} + {\hat C}_\alpha \pm \cos{2 \theta_{13}} ) } \sin^2{\frac{1}{2}(1+{\hat A}\mp {\hat C}_\alpha) \Delta_L} 
~.
\eeq

\end{widetext}
The lower sign applies above the atmospheric resonance, and the upper sign applies below it.  The term $P_3$, of order $\alpha^2$, is a correction term that improves the overall accuracy. To obtain these results, Freund linearized the oscillating terms over $\alpha$, the solar mass-squared difference.  His  results thus apply only  for baselines where 
\beq
\label{dnlim}
\alpha \Delta_L < 1 ~.
\eeq

In addition to observing  that his  expressions for the $\alpha$-expanded  eigenvalues are not valid for $|{\hat A} |  <  \alpha$, Freund~\cite{f}  notes a shortcoming of the  technique he uses when when there are resonances. The advantage of this technique is the possibility of  extracting  the mixing angles, as modified by the medium,  from the unitary transformation that diagonalizes the Hamiltonian.  

However,  in order to construct this transformation, it is necessary to  keep track of the ordering of the eigen${\it vectors}$ across resonances that characterize the transition  of interest. In the presence of resonances, the eigenvectors must be exchanged at their position  to preserve the proper ordering. Unavoidably, doing so  introduces discontinuities.  This is the source of the flip of sign   at at the location of the atmospheric resonance, ${\hat A} = \cos{2\theta_{13}}$, in Eq.~(\ref{expopal}).  This discontinuity is the price one pays to use this method.   We note in passing that these discontinuities are avoided in our Hamiltonian approach~\cite{jhk0} by using the Lagrange interpolation formula.

Freund examines the accuracy of the result given in Eq.~(\ref{expopal})  by comparing  to exact numerical simulations for a specific baseline and range of energies encompassing the atmospheric resonance.  From these results, given   in his Fig.~1, he concludes that the loss of precision of his full $\alpha$-expanded 
is less serious for small values of $\theta_{13}$, such as the value $\theta_{13} \approx 0.1$, which is the CHOOZ bound.  Unfortunately, the recent result Daya Bay~\cite{DB} gives a larger result, $\sin{\theta_{13}} \approx 0.15$, which makes the use of Freund's full result somewhat awkward.

\subsection{ Patched Result of Freund  }

Freund identifies the source of the inaccuracies to be the sub-leading terms in $\theta_{13}$ within his representation. Based on his understanding, Freund proposes a ``patched" result that he argues would be more appropriate than his full result  when it fails.
This leads to his ``patched" expressions,

\begin{widetext} 

 \beq
\label{expopall}
P^{e\mu}_0 ( \Delta_L,{\hat A} ) &=&  \sin^2{\theta_{23}} \frac{\sin^2{2 \theta_{13} }}{  ( {\hat A} - 1)^2   }  \sin^2{ ( {\hat A }- 1)\Delta_L }  \nonumber \\
P^{e\mu}_{\sin\delta}(\Delta_L,{\hat A}) &=& \alpha \sin{\delta_{cp} }  \frac{ \cos{\theta_{13}} \sin{2 \theta_{12}} \sin{2 \theta_{13}} 
\sin{2 \theta_{23}} } { {\hat A}( {\hat A} - 1) }  \sin{ \Delta_L }  \sin{{ \hat A } \Delta_L } \sin  {({\hat A} - 1)  \Delta_L }  \nonumber \\
P^{e\mu}_{\cos\delta}(\Delta_L,{\hat A}) &=&  \alpha \cos{\delta_{cp} } \frac{ \cos{\theta_{13}} \sin{2 \theta_{12}} \sin{2 \theta_{13}} 
\sin{2 \theta_{23}} } { {\hat A}( {\hat A} - 1) }  \cos{ \Delta_L }  \sin {{ \hat A } \Delta_L } \sin  {({\hat A} - 1)  \Delta_L }  \nonumber \\
P^{e\mu}_3(\Delta_L,{\hat A}) &=& \alpha^2 \frac{ \cos^2{\theta_{23}} \sin^2{2 \theta_{12}} } { {\hat A}^2 } \sin^2{{\hat A} \Delta_L}    ~,
\eeq

\end{widetext}
which appear in Eq.~(38a-38d) of Ref.~\cite{f}.  Freund notes that the $\theta_{13}$ dependence  is improved by retaining $ P^{e\mu}_3(\Delta_L,{\hat A})$.  He neglects $ P^{e\mu}_1(\Delta_L,{\hat A}) $ and $ P^{e\mu}_2(\Delta_L,{\hat A}) $ because he finds them much smaller than  $P^{e\mu}_{\sin\delta} (\Delta_L,{\hat A}) $ and $P^{e\mu}_{\cos\delta} (\Delta_L,{\hat A}) $.

\subsection{ Full Result of AHLO   }
\label{ahlot}

The AHLO neutrino oscillation probability,
\beq
\mathcal{P}(\nu_a  \rightarrow \nu_b) &\equiv& P^{\nu_a \to \nu_b} \nonumber \\
&=& | S^{ab} |^2 ~,
\eeq
is determined from an expansion of time-evolution operator $S^{ab}$ in $\sin^2\theta_{13}$.  The approach was originally proposed in Refs~\cite{ahlo,JO}, and details of its implementation are given in Appendix A of In Ref.~\cite{ahlo}.  Their final result is a first-order perturbation theory expansion of $S^{ab}$  in
$\sin^2\theta_{13}$, denoted there $S^\prime (t,t_0) $ and is equivalent to  a first-order  Taylor expansion of the ${\it entire}$ $\mathcal{P}(\nu_i  \rightarrow \nu_f) $ about $\xi_i = \sin^2\theta_{13} = 0$,
\beq
\label{OPFAHLO}
\mathcal{P}(\nu_a  \rightarrow \nu_b) &\approx& P^{ab}|_{\xi_i= \sin^2\theta_{13} =0} \nonumber \\
&+& \sin^2\theta_{13} \frac{ d P^{ab}}{d\xi_i}|_{\xi_i =0}  ~.
\eeq
Because this expansion involves the ${\it  total}$  derivative, $d P^{ab}/ d \xi$, the expanded eigenvalues are not used explicitly~\cite{ahlo,JO}.

The expression for $S^\prime (t,t_0) $ is found from the solution of the time-dependent Schroedinger equation given in Eq.~(A.9) of Ref.~\cite{ahlo},
\beq
\label{tdse1}
i\frac{d}{dt} S_0^\prime (t,t_0) &=& H_0^\prime S_0^\prime (t,t_0) ~,
\eeq
with the Hamiltonian $H_0^\prime$, expressed in the neutrino mass basis given in Eq.~(A.6),
\beq
\label{fulh}
 H_0^\prime &=& \frac{\Delta}{2}  \nonumber \\
&\times& \left( \begin{array}{cc} -\alpha \cos{2 \theta_{12}} + \hat A   & \alpha \sin{2 \theta_{12}} \\  \alpha \sin{2 \theta_{12}} & \alpha \cos{2 \theta_{12}} - \hat A  \end{array} \right ) ~,
\eeq
where
\beq
\Delta &\equiv& \frac{ m^2_3-m^2_1}{2 E}.
\eeq

The solution of Eq.~(\ref{tdse1}), Eq.~(22) of Ref.~\cite{ahlo},
\beq
S_0^\prime(t,t_0) &=& \left( \begin{array}{cc} \alpha(t,t_0) & \beta(t,t_0) \\  -\beta^*(t,t_0) & \alpha^*(t,t_0) \end{array} \right ) ~,
\eeq
is expressed in terms of  $\alpha(t,t_0)$ and $ \beta(t,t_0)$.  The quantities $\alpha(t,t_0)$ and $ \beta(t,t_0)$, which may be found in closed form,
determine  $ S^{ab} $ to leading order in $\sin\theta_{13}$,
\beq
S^\prime(t,t_0) &=& \left( \begin{array}{cc} \alpha & S_{12} \\ S_{21} & S_{22} \end{array} \right ) ~.
\eeq
All higher-order corrections are contained in the transformation  that rotates $ S_0^\prime$ back to the original flavor basis.

Defining the baseline $L$ in terms of the time $t-t_0$ for the neutrinos to travel between their production target and detector in  a uniform medium, as in Eq.~(\ref{Ldef}), all quantities become functions of $L$.  In particular,
$ S_{12}(t,t_0) \to   S_{12}(L) $, where
\beq
S_{12}(L) &=& \cos\theta_{23}\beta(L) - i\sin\theta_{23} A(L) ~.
\eeq
The quantity $A(L)= a A_a(L) + b A_b(L)$ is given as the first expression in Eq.~(A.14) of Ref.~\cite{ahlo}, with $A_a(L)$ defined in Eq.~(A.16) and $A_b(L) $ in Eq.~(A.17).  These quantities may be calculated analytically as integrals over $L$ involving $ f(L)$,
\beq
f(L) &=& e^{-i \bar\Delta L } ~,
\eeq
$\alpha(L)$, and $\beta(L)$.  
We have solved these equations for $S^\prime(t,t_0) $ exactly, and all numerical AHLO results are obtained from these expressions.  These resulting expressions  are, however,  is very complicated and too awkward to use in practice.  We have used them because our only interest here is to benchmark Refs~\cite{ahlo,JO}.
Simplified expressions for the matrix element  $ S_{12} $, which  describes $\nu_\mu \to \nu_e $ transitions, were used in Refs.~\cite{hjk1,khj2,hjk1a,khj1E}. Because these expressions entail a yet further set of approximations, for the purpose of this paper it serves no purpose to examine these simplified expressions.

Below, we numerically compare the oscillation probability obtained from $S^\prime (t,t_0)  $ to the exact one obtained from our Hamiltonian formulation.
The comparison is made  for the specific case of $e \to \mu$ transitions.

\section{ Assessing the  $\xi_i$-Expanded  $\mathcal{P}(\nu_i  \rightarrow \nu_f)$  }
\label{ExpComp}

In Sect.~\ref{NCall}, we compare the  $\alpha$-expanded oscillation probabilities, the  $\sin^2\theta_{13}$-expanded oscillation probabilities, and the exact oscillation probabilities.   From these results, we are able to identify the regions where both  $\xi_i$-expanded oscillation probabilities might be improved by using our Hamiltonian formulation.

\subsection{Regions of $\hat A$ }

We assess the expanded oscillation probabilities over specific regions of  ${\hat A}$ and $\Delta_L$.  For this purpose,  the interval $\hat A > -0.8$,  broken down  into five sub-regions spanning the interval $\hat A_i < \hat A < \hat A_f$.  

The first region is the solar resonance region containing the solar resonance at $\hat A = \alpha$.  This region extends from $\hat A_i = 0$  to  $\hat A_f = 0.1$.  Motivated by the representation of the $\sin^2\theta_{13}$-expanded eigenvalues given in Appendix~\ref{diagH}, we split the solar resonance region into  the deep solar resonance region, $0< \hat A <  \alpha $ and the far solar resonance region,  $\alpha < \hat A <  0.1 $.

The second region, the transition region, extends from  $\hat A_i = 0.1$  to  $\hat A_f = \hat A_2$, where  $ A_2 \approx 0.538 $ is the value of $\hat A$ at which the periods of the two most slowly varying Bessel functions,  $j_0( {\hat \Delta}[1] )$ and $j_0( {\hat \Delta}[3] )$, are equal.  The third region includes the atmospheric resonance  at $\hat A = \hat A_0 \approx 1$ and  extends from $\hat A_i = \hat A_2$,  $\hat A_f = 1.2$,  The asymptotic  region includes all values of  $\hat A  > 1.2 $.   Finally, we consider the region extending from $\hat A_i = -0.8 $ to  $\hat A_f = +0.8$.

Within each region considered, $\Delta_L$ is chosen so that $\mathcal{P}(\nu_e  \rightarrow \nu_\mu)$  displays the greatest sensitivity to approximations.  This is assured by requiring that each of the three Bessel functions constituting  the oscillation probability  are of comparable size and interfere both constructively and destructively.  For maximal interference,  each  of  these Bessel functions   $j_0(\Delta_\ell)$ should undergo at least one period of oscillation when the neutrino is detected at a distance $\Delta_L$ from the source.  This is fully discussed in Ref.~\cite{jhk0}.  

The  maximal interference within the solar resonance region was determined to occur for $\Delta_{L} = 60$;  the transition  region for $\Delta_{L} = 17$; the  atmospheric resonance  region, for $\Delta_L = 4$; the  asymptotic region for  $\Delta_L = 4$. For the region extending from $\hat A_i = -0.8 $ to  $\hat A_f = +0.8$,  $\Delta_L = 35$.

The $\Delta_L$ dependence is examined for two values of $\hat A$, one in the solar resonance region and one in the atmospheric resonance region. 
In all cases, the oscillation probability is examined for $\delta_{cp} = \pi/4$,  with the other parameters taken from the SNM. 

\subsection{ Extrapolations}

The results shown in this paper  may be extrapolated to a variety of values baselines, medium properties, and neutrino energies within the regions under using Eqs.~(\ref{Deldeff},\ref{ahatrho}),
\beq
\label{Deldefalt3}
L[{\rm Km}] &=& 5.04 \times 10^3 \frac{ |{\hat A}| \Delta_L }{ R \rho[{\rm gm/cm}^3] } \nonumber \\
E[{\rm GeV}] &=& 15.3 \frac{ |{\hat A}|}{ R \rho[{\rm gm/cm}^3] } ~,
\eeq
for a medium described by $\rho$ and $R$.  Using Eq.~(\ref{Deldefalt3}), the oscillation probability may be extrapolated  to a variety of values of neutrino energy, base lines, and medium properties once  it is known for a specific set of values $(\Delta_L,\hat A)$

\section{Comparison  of  the Full and Patched $\mathcal{P}(\nu_e  \rightarrow \nu_\mu)$ of Freund to the  Full  $\mathcal{P}(\nu_e  \rightarrow \nu_\mu)$ of AHLO} 

\label{NCall}

In this section,  we assess simplified expressions  for $\mathcal{P}(\nu_e  \rightarrow \nu_\mu)$ found in Refs~\cite{f,ahlo,JO}.   These expressions are the only published attempts made to simplify the oscillation probability.

We compare the simplified $\mathcal{P}(\nu_e  \rightarrow \nu_\mu)$  to  the corresponding exact  oscillation probability~\cite{jhk0} evaluated with the exact eigenvalues to determine the relative and absolute accuracy of the approximate oscillation probabilities commonly used for the  analysis and prediction of neutrino oscillation phenomena. From these results, one easily identifies which of the $\xi_i$-expanded oscillation probabilities best describes each region.

Freund's patched is shown as a short-dashed curve, his full result as a  medium-dashed curve, the AHLO oscillation probability as a dot-dashed curve, and the oscillation probability of  our Hamiltonian formulation~\cite{jhk0} evaluated with the exact eigenvalues as a solid curve.


\subsection{ $\Delta_L$ Dependence of the $\xi_i$-Expanded $\mathcal{P}(\nu_e  \rightarrow \nu_\mu)$ }
\label{freuPD}

In this section, we compare the oscillation probabilities over intervals of $\Delta_L$ for several values of ${\hat A}$, one near the solar resonance and the other  near the atmospheric resonance.

\subsubsection{Near the Solar Resonance}

\begin{figure}
\centerline{\epsfig{file=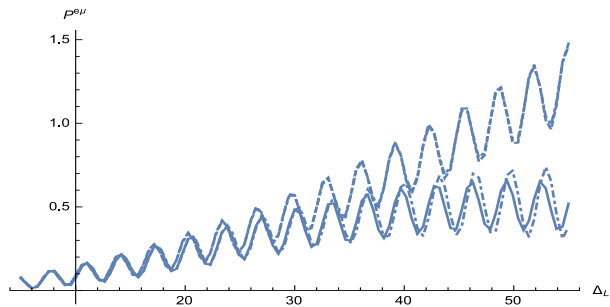,height=4cm,width=8.cm}}
\caption{$P^{e\mu}(\Delta_L,{\hat A})$ in the solar-resonance region (${\hat A}=0.0102$) over the interval $5<\Delta_L<55$ for neutrinos in matter. Parameters are taken from the SNM. Exact result (solid curve).  Patched $\alpha$-expanded result of Freund~\cite{f} (short-dashed curve).  Full  $\alpha$-expanded result of Freund~\cite{f} (medium-dashed curve). Full $\sin^2\theta_{13}$ result of AHLO~\cite{ahlo}   (dot-dashed curve)}
\label{figcptSec1}
\end{figure}

We first look at the $\Delta_L$-dependence of $P^{e\to \mu}(\Delta_L,{\hat A})$ in the solar resonance region for $\hat A = 0.0102$. Figure~\ref{figcptSec1} shows that Freund's full $\alpha$-expanded oscillation probability and his patched oscillation probability are both in reasonable agreement with the exact result for $\Delta_L < 20$, at which point both of Freund's oscillation probabilities begin to depart from the exact 
oscillation probability.  They increasingly diverge from the exact result as  $\Delta_L$ increases, which reflects the breakdown of the $\alpha$-expansion in the solar resonance region. On the other hand, the AHLO result agrees with the exact result comparatively well over the entire interval of $\Delta_L$ shown.

\subsubsection{Near the Atmospheric Resonance}

\begin{figure}
\centerline{\epsfig{file=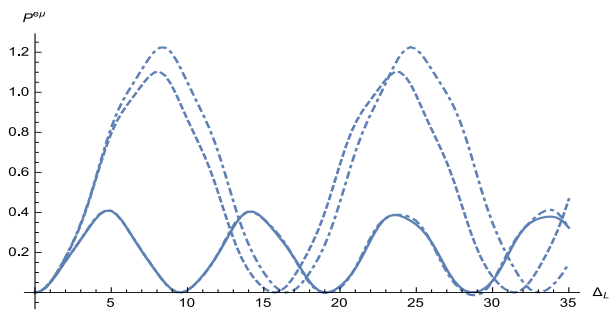,height=4cm,width=8.cm}}
\caption{ Freund $P^{e\mu}(\Delta_L,{\hat A})$ in the atmospheric resonance region (${\hat A}=0.8$) over the interval $0<\Delta_L<35$ for neutrinos in matter. Parameters are taken from the SNM. Exact result (solid curve).  Patched $\alpha$-expanded result of Freund~\cite{f} (short-dashed curve).  Full  $\alpha$-expanded result of Freund~\cite{f} (medium-dashed curve). Full $\sin^2\theta_{13}$ result of AHLO~\cite{ahlo}   (dot-dashed curve)}
\label{figcptSec2}
\end{figure}

We next look at the $\Delta_L$-dependence of $P^{e\to \mu}(\Delta_L,{\hat A})$ in the atmospheric  resonance region for $\hat A = 0.8$, a value of ${\hat A}$ below the atmospheric resonance. Figure~\ref{figcptSec2} shows that Freund's full $\alpha$-expanded oscillation probability is in reasonable agreement with the exact result over the entire interval of $\Delta_L$ shown.  On the other hand, Freund's patched $\alpha$-expanded oscillation probability and the AHLO result are quite similar but are in substantial disagreement  with the exact result over the same interval of $\Delta_L$.

\subsection{ The $\hat A$ Dependence of the $\xi_i$-Expanded $\mathcal{P}(\nu_i  \rightarrow \nu_f)$  }
\label{freuPA}

Below, we examine the $\hat A$ dependence over the five regions defined earlier.

\subsubsection{Solar Resonance Region:   $0< \hat A < 0.1$ }

\begin{figure}
\centerline{\epsfig{file=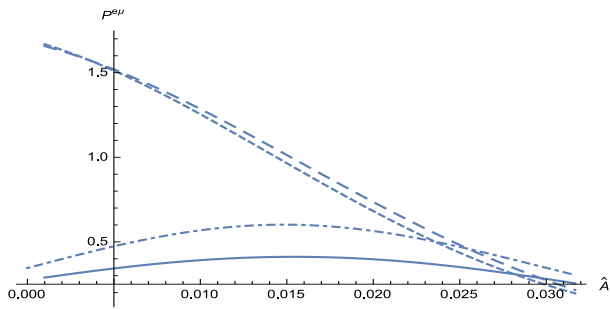,height=4cm,width=8.cm}}
\caption{$P^{e\mu}(\Delta_L,{\hat A})$ over the deep solar resonance region, $0<\hat A< \alpha$, with $\Delta_L=60$ for neutrinos in matter. Parameters are taken from the SNM. Exact result (solid curve).  Patched $\alpha$-expanded result of Freund~\cite{f} (short-dashed curve).  Full  $\alpha$-expanded result of Freund~\cite{f} (medium-dashed curve). Full $\sin^2\theta_{13}$ result of AHLO~\cite{ahlo}   (dot-dashed curve)}
\label{figcptSec3}
\end{figure}

\begin{figure}
\centerline{\epsfig{file=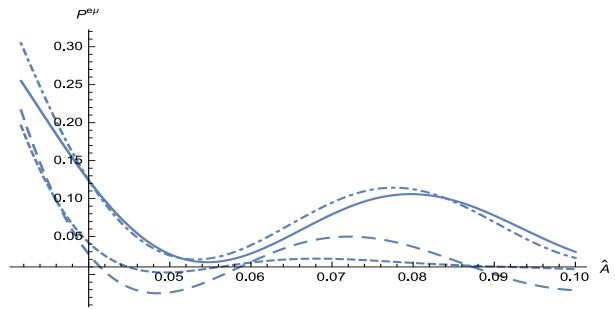,height=4cm,width=8.cm}}
\caption{$P^{e\mu}(\Delta_L,{\hat A})$ over the far solar resonance region,  $\alpha<\hat A<0.1$ with $\Delta_L=60$ for neutrinos in matter. Parameters are taken from the SNM. Exact result (solid curve).  Patched $\alpha$-expanded result of Freund~\cite{f} (short-dashed curve).  Full  $\alpha$-expanded result of Freund~\cite{f} (medium-dashed curve). Full $\sin^2\theta_{13}$ result of AHLO~\cite{ahlo}   (dot-dashed curve)}
\label{figcptSec4}
\end{figure}

The deep solar resonance region, which extends over the interval  $0< \hat A <  \alpha $, is shown in Fig.~\ref{figcptSec3}.  The far solar resonance region, which extends over the interval $\alpha < \hat A <  0.1 $, is shown  in Fig.~\ref{figcptSec4}.  The results of Freund~\cite{f}  and AHLO~\cite{ahlo} are compared to the exact result within these two regions.  The $\alpha$-expansion is known to fail within the solar resonance region.~\cite{jhk0} because one of the eigenvalue branch point lies at $\alpha = 0$. 

Although Freund asserts that his results fails only below the  solar resonance, $\hat A < \alpha$, it is clear from Fig.~\ref{figcptSec4} that  the $\alpha$-expanded results are poor approximation out to $\hat A \approx 0.1$.  The failure becomes increasingly severe as $\alpha \to 0$.  This is evident for both Freund's patched and full results.

Clearly, the  full $\sin^2\theta_{13}$-expanded result of AHLO does much better within the solar resonance region. However, Fig.~\ref{figcptSec3} shows that the AHLO result does not approach the exact result in the limit $\alpha \to 0$.

\subsubsection{ $0.1 < \hat A < \hat A_2$ }

\begin{figure}
\centerline{\epsfig{file=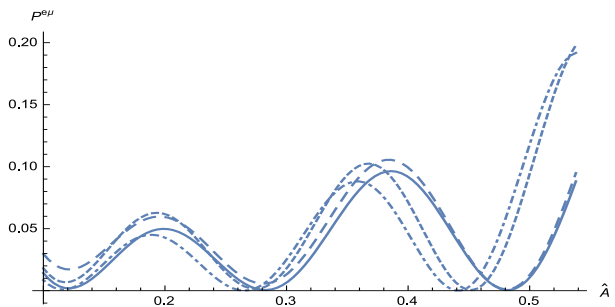,,height=4cm,width=8.cm}}
\caption{ $P^{e\mu}(\Delta_L,{\hat A})$ for $\Delta_L=17$ over the over the transition region,  $0.1<{\hat A}< \hat A_2$, for neutrinos in matter. Parameters are taken from the SNM.   Exact result (solid curve).  Patched $\alpha$-expanded result of Freund~\cite{f} (short-dashed curve).  Full  $\alpha$-expanded result of Freund~\cite{f} (medium-dashed curve). Full $\sin^2\theta_{13}$ result of AHLO~\cite{ahlo}   (dot-dashed curve)}
\label{figcptSec5}
\end{figure}

In Fig.~\ref{figcptSec5}, we compare the oscillation probabilities within the transition region. Comparing the short-dashed and  medium-dashed curves to the the solid curve, it is clear that that Freund's patched and full results agree well with the exact result for $0.1 < \hat A < 0.3$, but for larger values of $\hat A$ his patched result increasingly departs from the exact result, whereas his full result tracks the exact result throughout the transition region.  

Comparing the  dot-dashed and solid curves, it is clear that that Freund's patched and the AHOL result provide rather similar descriptions of the oscillation probability throughout the transition region.

\subsubsection{ $\hat A_2 < \hat A < 1.2$ }

\begin{figure}
\centerline{\epsfig{file=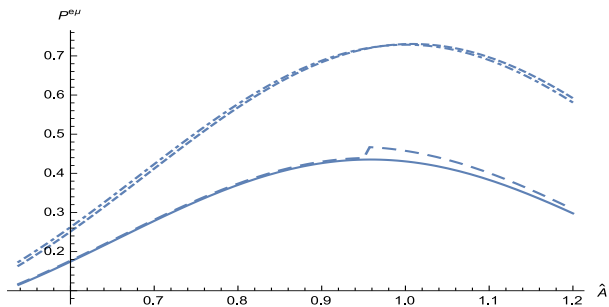,height=4cm,width=8.cm}}
\caption{ $P^{e\mu}(\Delta_L,{\hat A})$ for $\Delta_L=4$ over the atmospheric resonance  region, $\hat A_2 <{\hat A}<1.2$, for neutrinos in matter. Parameters are taken from the SNM.   Exact result (solid curve).  Patched $\alpha$-expanded result of Freund~\cite{f} (short-dashed curve).  Full  $\alpha$-expanded result of Freund~\cite{f} (medium-dashed curve). Full $\sin^2\theta_{13}$ result of AHLO~\cite{ahlo}   (dot-dashed curve)}
\label{figcptSec6}
\end{figure}

In Fig.~\ref{figcptSec6}, we compare the oscillation probabilities over the atmospheric resonance  region.  Comparing the dot-dashed and medium-dashed curves, it is clear that that the departure of Freund's full result from his patched result continues to increase across the atmospheric resonance region. Comparing the  dot-dashed and solid curves, it is clear that that Freund's patched and the AHOL result continue to provide rather similar descriptions of the oscillation probability throughout the atmospheric resonance region. 

Freund's full result provides a reasonable description of the exact result below the atmospheric resonance, at which point  his result is discontinuous and lies above the exact result for larger values of $\hat A$.

\subsubsection{ $1.2 < \hat A < 2.5$ }

\begin{figure}
\centerline{\epsfig{file=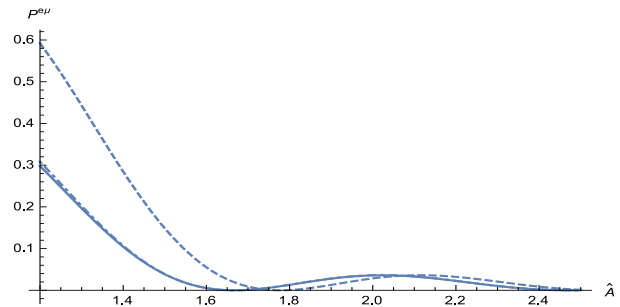,height=4cm,width=8.cm}}
\caption{ $P^{e\mu}(\Delta_L,{\hat A})$ {\it vs} ${\hat A}$  for $\Delta_L=4$ within the asymptotic  region, $1.2 < {\hat A} < 2.5$ for neutrinos in matter.
 Parameters are taken from the SNM.    Exact result (solid curve).  Patched $\alpha$-expanded result of Freund~\cite{f} (short-dashed curve).  Full  $\alpha$-expanded result of Freund~\cite{f} (medium-dashed curve). Full $\sin^2\theta_{13}$ result of AHLO~\cite{ahlo}   (dot-dashed curve)}
\label{figcptSec7}
\end{figure}

In Fig.~\ref{figcptSec7}, we compare the oscillation probabilities over a portion of the asymptotic region.  Comparing the dot-dashed and short-dashed curves, we see once again  the departure of Freund's patched result from his patched result.  However, this departure appears to   decreases above the atmospheric resonance region  and approaches the exact result for the larger values of $\hat A$.

\subsubsection{ $-0.8 < \hat A < 0.8$ }

\begin{figure}
\centerline{\epsfig{file=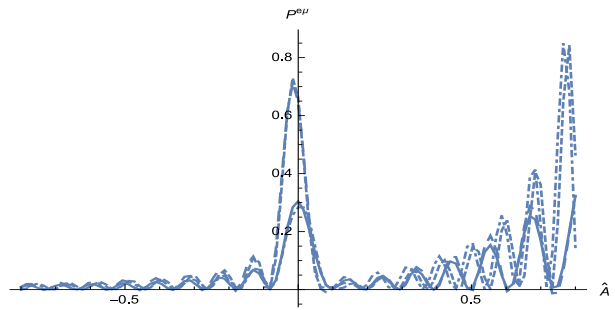,height=4cm,width=8.cm}}
\caption{ $P^{e\mu}(\Delta_L,{\hat A})$ over the interval $-0.8<\hat A<0.8$ with $\Delta_L=35$ for neutrinos in matter. Parameters are taken from the SNM.  Exact result (solid curve).  Patched $\alpha$-expanded result of Freund~\cite{f} (short-dashed curve).  Full  $\alpha$-expanded result of Freund~\cite{f} (medium-dashed curve). Full $\sin^2\theta_{13}$ result of AHLO~\cite{ahlo}   (dot-dashed curve)}
\label{figcptSec8}
\end{figure}

 Because the branch point of the $\alpha$ expansion occurs for  ${\hat A} = 0$, it will adversely impact the total oscillation probability for both neutrinos and anti-neutrinos  This is confirmed in Fig.~\ref{figcptSec8}.

\subsection{Discussion}

In this section, we calibrated the  accuracy of Freund's~\cite{f}  full and patched exact oscillation probabilities  and the full AHLO~\cite{ahlo}  oscillation probability.  We were able to identify which of the approximate solutions best described the exact result within and above the solar resonance region.  Few studies of this nature have ever been made.

Freund's patched result  was found to be  significantly less accurate than the full result in all  regions. Freund's  full  result was  calculated and compared  to the exact result in all  regions, including the solar resonance region.    The failure of Freund's full result within the solar resonance region is evident from Figs.~\ref{figcptSec3},\ref{figcptSec4}; we also confirmed that  the $\sin^2\theta_{13}$ expansion leads to inaccuracy  in the  vicinity of the atmospheric resonance.  Both  inaccuracies grow as $\hat A$ approaches the  branch points.  

In the vicinity of  the atmospheric resonance, $\hat A \approx  \cos 2\theta_{13}$, we found that neither the  AHLO result nor the full result of Freund provides a very good description of the oscillation probability. This is expected for the  $\sin^2\theta_{13}$ expansion because the corresponding branch point lies close to the atmospheric resonance. Freund's full result is discontinuous at the position of the atmospheric resonance.  This discontinuity is quite noticeable when the Daya Bay value of $\theta_{13}$ is adopted.

It was clear from our calculations that our Hamiltonian formulation needs no patch to avoid any discontinuity such as that present in Freund's full result.  

Freund, in his paper~\cite{f}, does actually compare his full and patched results to an exact calculation of the oscillation probability [his Fig.~(1)].  His  results, when viewed casually, may appear inconsistent with those given here, especially with regard to the theoretical error of his patched result.  

However, one should note that Freund's comparisons are presented on a logarithmic plot of multiple decades, whereas ours are given on a linear plot.  On a logarithmic  scale, factors of two are often get lost in the noise. Additionally, Freund considers a broad range of model parameters, which makes it difficult to pinpoint the portion of his presentation that reflects the SNM, which is of course the model underlying our analysis.

We also found that the $\sin^2\theta_{13}$ expansion~\cite{ahlo,JO}  represents of the exact oscillation probability relatively accurately within the solar resonance region, but its accuracy in other regions is comparable to Freund's patched result.  We also found that in the region between the solar and atmospheric resonances, $0.1 < {\hat A}  < \hat A_2 $, the expansion in $\alpha$ is  a good representation of the exact result, whereas the expansion in $\sin^2 \theta_{13}$ rapidly deteriorates with increasing ${\hat A}$.

In the vicinity of  the atmospheric resonance, $\hat A \approx  \cos 2\theta_{13}$, we found that neither the  AHLO result nor the full Freund result provides a very good description of the oscillation probability. This is expected for the  $\sin^2\theta_{13}$ expansion because the corresponding branch point lies close to the atmospheric resonance.  A careful reading of Ref.~\cite{f} reveals that some inaccuracy of the $\alpha$ expansion would have been anticipated there in light of the Daya Bay result~\cite{DB}, but its extent may be a surprising disappointment to  some.  The ${\it asymptotic}$ region, ${\hat A}  > 2.0$ is shown in  Fig.~\ref{figcptSec7}.  

Even though  expanded eigenvalues are not used explicitly for the  $\sin^2\theta_{13}$-expanded oscillation probability, Eq.~(\ref{OPFAHLO}) is still a poor representation of the exact result in the vicinity of the branch point at ${\hat A } = {\hat A}_0 $, where the eigenvalue expansion fails.
To separate the  theoretical errors into those arising from the expansion of the oscillation probability from those arising from  the expansion of the eigenvalues we have relied  on the  numerical techniques.

\section{Approximate Formulation of $\mathcal{P}(\nu_i  \rightarrow \nu_f)$ Accurate to a Few Percent }
\label{Xcompare}

Because subtle but important trends are more easily identified by examining analytic expressions than by scanning through numerical lists, simpler  and most accurate analytic expressions would be  significant considering that higher quality data are expected to become available at  future neutrino facilities.  In our original publication~\cite{jhk0}, we compare  exact oscillation probabilities to those  found from our Hamiltonian formulation evaluated with $\xi$-expanded eigenvalues.

\subsection{ Above the Solar Resonance Region, $\hat A > 0.1$}
\label{XcompareASRR}

In our original publication~\cite{jhk0}, we find that the errors of $O(\alpha^2)$ expected in regions  sufficiently remote from the branch point of the $\alpha$-expanded eigenvalues obtain for $\hat A > 0.1$. From this we estimate that simple analytic expressions for the oscillation probability accurate to a few percent are quite likely over this region using  our Hamiltonian formulation~\cite{jhk0}.

To identify the regions where  expressions for the oscillation probability of our Hamiltonian formulation~\cite{jhk0} are improvements over simplified oscillation probabilities found by Freund.~\cite{f}, we need to examine the accuracy of Freund's full and patched oscillation probabilities.  Figures presented in Sect.~\ref{NCall} assess this accuracy.   Because the accuracy of Freund's patched oscillation probabilities is found to be vastly inferior to that of his full oscillation probability, his patched oscillation probabilities are not further considered.  

The full  $\alpha$-expanded oscillation probability of Freund~\cite{f} is compared to  the exact oscillation probability over the interval $0.1 < \hat A < \hat A_2$ in  Fig.~\ref{figcptSec5} of the present paper.  We see from this figure  that  Freund's full  oscillation probability~\cite{f} becomes increasingly less accurate as $\hat A \to 0.1$. On the other hand, the oscillation probability of our Hamiltonian formulation evaluated with $\alpha$-expanded eigenvalues, which is compared to  the exact oscillation probability over the same interval in Fig.~4 of Ref.~\cite{jhk0}, remains quite accurate as $\hat A \to 0.1$. We conclude from these results that simple expressions for the oscillation probability more accurate than Freund's full result over this interval may be found using the results our Hamiltonian formulation.

Freund's full  oscillation probability ~\cite{f} is  compared to  the exact oscillation probability in the vicinity of the atmospheric resonance in  Fig.~\ref{figcptSec6}
of the present paper.  We see there  that  Freund's full  oscillation probability~\cite{f} is discontinuous at the location of the atmospheric resonance, whereas the oscillation probability  obtained from our Hamiltonian formulation evaluated with $\alpha$-expanded eigenvalues given in Fig.~6 of our original paper~\cite{jhk0}  is continuous throughout the atmospheric resonance region.  Because it is also  a very accurate representation of the exact result, we conclude that simple expressions for the oscillation probability that are more accurate than Freund's full result in the vicinity of the atmospheric resonance may be found using the results our Hamiltonian formulation. 

The full  $\alpha$-expanded oscillation probability of Freund~\cite{f} is compared to  the exact oscillation probability for asymptotic values of $\hat A$ in  Fig.~\ref{figcptSec7} of the present paper.  We see from this figure and  Fig.~7 of our original paper~\cite{jhk0} that Freund's full oscillation probability~\cite{f} and that obtained from our Hamiltonian formulation evaluated with $\alpha$-expanded eigenvalues are of comparable accuracy for asymptotic values of $\hat A$.   We conclude that simple expressions for the oscillation probability of accuracy comparable to Freund's full result   may be found using the results our Hamiltonian formulation for asymptotic values of $\hat A$. 

In our original publication~\cite{jhk0}, we also compare  exact oscillation probabilities to those  found from our Hamiltonian formulation evaluated with $\sin^2\theta_{13}$-expanded eigenvalues.  The accuracy with which the oscillation probability of our Hamiltonian formulation evaluated with 
$\sin^2\theta_{13}$-expanded eigenvalues reproduces the exact oscillation probability is strikingly good over the region $0.1 < \hat A <  \hat A_2$, as  seen in Fig.~4 of our original paper.  This establishes that simple and accurate analytic expressions for the oscillation probability are also likely to be found  somewhat above the solar resonance region  using results of our Hamiltonian formulation~\cite{jhk0} with $\sin^2\theta_{13}$-expanded eigenvalues.  

\subsection{ Solar Resonance Region, $0 < \hat A < 0.1$}
\label{XcompareSRR}

Similar considerations apply to $\sin^2\theta_{13}$-expanded oscillation probabilities.  Close agreement between the exact oscillation probability and those of our Hamiltonian formulation evaluated with  $\sin^2\theta_{13}$-expanded eigenvalues is seen over the region $0 < \hat A < 0.1$ seen in Fig.~3 of our original paper~\cite{jhk0}.  This suggests that simpler  and most accurate analytic expressions for the oscillation probability within the solar resonance region are likely to be found  using results obtained from our Hamiltonian formulation~\cite{jhk0}.  

The figures in our original paper~\cite{jhk0} also  show that for values of $\hat A$ closer to the branch point of the $\sin^2\theta_{13}$-expanded eigenvalues, the accuracy of the oscillation probabilities deteriorate rather rapidly.

AS before, to identify the regions where  expressions for the oscillation probability obtained from our Hamiltonian formulation~\cite{jhk0} are improvements over simplified oscillation probabilities found by AHLO~\cite{ahlo},  we need to examine the accuracy of the full AHLO oscillation probabilities.  Figures presented in Sect.~\ref{NCall} assess this accuracy.

We see from the results appearing in Figs.~\ref{figcptSec3} to Fig.~\ref{figcptSec7}  of the present paper that  the full  $\sin^2\theta_{13}$-expanded oscillation probability of AHLO agrees rather poorly with the exact result over all regions except for  the far solar resonance region.  Within the far solar resonance region, the full  $\sin^2\theta_{13}$-expanded oscillation probability of our Hamiltonian formulation~\cite{ahlo} is reasonably accurate.

From  these results, we conclude that simple expressions of accuracy greater than a few percent are likely to be found from the results our Hamiltonian formulation  evaluated with $\sin^2\theta_{13}$-expanded eigenvalues. Such oscillation probabilities would then be vastly more accurate than those of AHLO~\cite{ahlo}.

\section{ Summary and Conclusions}

In this paper, we assessed  the theoretical errors of simplified  neutrino oscillation probabilities obtained from  various  approximate formulations. The errors  of the simplified oscillation probabilities obtained in  Refs.~\cite{f,ahlo,JO} are associated, in large part, with branch points~\cite{jhk0}  of the medium-modified neutrino eigenvalues.   Currently, there is essentially no  understanding  of the errors of these simplified oscillation probabilities.    We also assessed the likelihood that simple analytical approximations to  three-flavor neutrino oscillation probabilities with accuracy of a few percent would be found using our recent Hamiltonian formulation~\cite{jhk0}.  We considered  flavor-changing transitions in the   $(\nu_e , \nu_\mu)$  sector within the Standard Neutrino Model, the only cases considered in  Refs.~\cite{f,ahlo,JO}.

We compared the approximate results over various regions of $\hat A$  to  identify which  gave the best description of the exact results.  From this comparison, we found that the full results of Freund are  satisfactory everywhere  except for the solar- and atmospheric-resonance regions.  Those of AHLO describe the solar-resonance region well, but they are complicated and rapidly deteriorate outside this region. The simple neutrino oscillation probabilities of the approximate formulation we propose  in Sect.~\ref{Xcompare} are accurate to a few percent errors within essentially all regions of interest.

We thus conclude from Sect.~\ref{Xcompare} that our Hamiltonian formulation offers the  opportunity to obtain  approximate oscillation probabilities vastly more accurate than those presently available. This result is significant because it would obviate the need for exact  computer simulations under many circumstances, thus simplifying  analysis and prediction at current and future and next-generation neutrino neutrino facilities.

Making use of the guidance provided by the results of this paper, in future work~\cite{jk1} we derive simple approximations with few percent errors within essentially all regions of interest using our Hamiltonian formulation.

\appendix

\section{Oscillation Probability in Our Hamiltonian Formulation }
\label{SLTH}

In this appendix,  we  give alternative expressions  for the  $\sin^2\theta_{13}$-expanded eigenvalues  equivalent to  those given in  Ref.~\cite{jhk0}.   Separate expressions  are given for $\hat A$ above and below $\hat A = \alpha$  in terms of the ratio  of the small parameters of the SNM,  
 $R_p \equiv \sin^2\theta_{13}/\alpha \approx 0.7$


\subsection{Alternative Representation for the  $ \sin^2\theta_{13}$-Expanded Eigenvalues }
\label{diagH}

We introduce  the notation $ {\hat {\bar E}}^L_{\theta \ell}$ to represent  the $\sin^2\theta_{13}$-expanded eigenvalues ${\hat {\bar E}}_\ell $   for  $\hat A < \alpha$, and $ {\hat {\bar E}}^G_{\theta \ell}$  to represent them for  $\hat A > \alpha$.  The representation of ${\hat {\bar E}}_\ell $  given in terms of ${\hat {\bar E}}^L_{\theta \ell}$ and ${\hat {\bar E}}^G_{\theta \ell}$ is equivalent to the representation ${\hat {\bar E}}_\ell $  given in Ref.~\cite{jhk0}.

 \subsubsection{ $ \hat A < \alpha$ }

The following expressions for  $ {\hat {\bar E}}^L_{\theta \ell}$ are found  by replacing $\hat A \to  \alpha R_s$ and  $ {\hat C}_\theta \to \alpha C_T$ in
 ${\hat {\bar E}}_\ell $  given in Ref.~\cite{jhk0},
\beq
\label{evL}
 {\hat {\bar E}}^L_{\theta 1} &=& \frac{\alpha }{2 } ( R_s  + 1  -   C_T  ) \nonumber \\ 
&+& \frac{   \alpha  R_p    R_s } { 2(1-y)   C_T } (2 -{\hat A}_0  - \alpha - \alpha  C_T)  \nonumber \\
&+&   \frac{  R_p  y } { 2  C_T} (2 - 4{\hat A}_0 + {\hat A}_0^2 -2 \alpha + 2 {\hat A}_0 \alpha) \nonumber \\
 {\hat {\bar E}}^L_{\theta 2}
&=& \frac{\alpha }{2 } (    R_s  + 1  +   C_T  ) \nonumber \\ 
&-&  \frac{ \alpha   R_p   R_s } { 2(1-y) C_T } (2 -{\hat A}_0  - \alpha + \alpha  C_T )  \nonumber \\
&-&  \frac{  R_p y} { 2   C_T } (2 - 4{\hat A}_0 + {\hat A}_0^2 -2 \alpha + 2 {\hat A}_0 \alpha)
\nonumber \\
{\hat {\bar E}}^L_{\theta 3}
&=& 1 +   \frac{ \alpha^2 R_p R_s }{1-y}  ~.
\eeq

 \subsubsection{ $ \hat A > \alpha$ }

The following expressions for  $ {\hat {\bar E}}^G_{\theta \ell}$ are found  by replacing $\hat A \to  \alpha /R_s$ and ${\hat C}_\theta \to \hat A   C_T$  in ${\hat {\bar E}}_\ell $  given in Ref.~\cite{jhk0},
\beq
\label{evG}
 {\hat {\bar E}}^G_{\theta 1} &=& \frac{\alpha }{2 R_s} ( R_s  + 1  -   C_T  ) \nonumber \\ 
&+&  \frac{    \alpha R_p  } { 2R_s(1-y)   C_T } (2 R_s  -{\hat A}_0 R_s  - \alpha R_s  - \alpha  C_T)  \nonumber \\
&+&  \frac{ y R_s  R_p } { 2  C_T} (2 - 4{\hat A}_0 + {\hat A}_0^2 -2 \alpha + 2 {\hat A}_0 \alpha) \nonumber \\
 {\hat {\bar E}}^G_{\theta 2}
&=& \frac{\alpha }{2 R_s} (    R_s  + 1  +  C_T  ) \nonumber \\ 
&-& \frac{   \alpha R_p    } { 2R_s (1-y) C_T } (2 R_s  -{\hat A}_0 R_s  - \alpha R_s  + \alpha  C_T )  \nonumber \\
&-&  \frac{y R_s  R_p  } { 2  C_T } (2 - 4{\hat A}_0 + {\hat A}_0^2 -2 \alpha + 2 {\hat A}_0 \alpha)
\nonumber \\
{\hat {\bar E}}^G_{\theta 3}
&=& 1 + \frac{ \alpha^2 R_p  }{R_s(1-y)}  ~.
\eeq

\subsubsection{ Discussion  }

Because Eqs.~(\ref{evL},\ref{evG}) are equivalent  representations of  the eigenvalues ${\hat {\bar E}}_\ell $    and their differences, they are accurate to a few percent for $\hat A < 0.6$. The following points are also worthy of noting. 

The same  quantity 
\beq
\label{CTdef}
C_T &\equiv&  \sqrt{1 + R_s^2 - 2 R_s \cos\theta_{12} } \nonumber  \\
&\approx& 1~,
\eeq
appears in the expressions for the eigenvalues for both  $\hat A < \alpha$ and $\hat A > \alpha$.

Above the branch point at $y=1$,  ${\hat {\bar E}}^L_{\theta \ell}$ and ${\hat {\bar E}}^G_{\theta \ell}$ are related:

(1) $ {\hat {\bar E}}_{\theta 2}  =  {\hat {\bar E}}_{\theta 1}  |_{  C_T  \to - C_T}$ and  

(2) ${\hat {\bar E}}_{\theta 3}$ and ${\hat {\bar E}}_{\theta 2}$ exchange roles.


\section{Exact Analytic Results for the Coefficients $ w^{(mn)}_i[\ell] $ }
\label{wcoef}

Full expressions for the coefficients $w_{i;n}^{(mn)}$ appearing in the expressions for the partial oscillation probabilities of Ref.~\cite{jhk0}  are given  in this appendix.  They are expressed in terms parameters of the SNM and the following functions of them,
\beq
\label{Cpmval2}
C_1^{(\pm)} &\equiv& \cos^2{ \theta_{12}} \cos^2{ \theta_{23}} \pm \sin^2{ \theta_{23}} \sin^2{ \theta_{12}}  \sin^2{ \theta_{13}} \nonumber \\
C_2^{(\pm)} &\equiv& \cos^2{ \theta_{12}} \pm \sin^2{ \theta_{12}}  \sin^2{ \theta_{13}} ~,
\eeq
where $C_2^{(\pm)}$ was defined earlier in Eq.~(\ref{Cbval}).  In the SNM, many of the  terms in $w_{i;n}^{(mn)}$ are quite small and may be dropped in practice.

\subsection{ Coefficients $ w_{cos^2}^{(mn)} $ }

\beq
w_{cos^2;0}^{(22)} &=&  \frac{{\hat A} \alpha^2}{4} \cos^2{ \theta_{13} } \sin^2{ 2\theta_{12} } \sin^2{ \theta_{13} } (1- \alpha \sin^2{ \theta_{12} } ) \nonumber \\
w_{cos^2;1}^{(22)} &=& - \frac{\alpha^3}{4} \sin^2{2 \theta_{12} } \sin^2{ \theta_{13} } \nonumber \\
w_{cos^2;2}^{(22)} &=& \frac{\alpha^2}{4} \sin^2{2 \theta_{12} } \sin^2{ \theta_{13} }
\eeq

\subsection{ Coefficients $ w_{cos}^{(mn)} $ }

\newpage

\begin{widetext}

\beq
w_{cos;0}^{(12)} &=&  - \frac{\alpha}{8} \cos{ \theta_{13} } \sin{2 \theta_{12} } \sin{ 2 \theta_{13} }  ( (1-\alpha)^2 - {\hat A} ( \cos{2\theta_{13} } +  \alpha  ( 1- 2\cos{2\theta_{13} } \sin^2 { \theta_{12} } ) \nonumber \\
&-& 2 \alpha^2 \sin^2{ \theta_{12} } 
C^{(+)}_2) ) \nonumber \\
w_{cos;1}^{(12)} &=& -\frac{\alpha}{8} \cos{ \theta_{13} } \sin{2 \theta_{12} } \sin{ 2\theta_{13} } 
( 1 + \alpha - 2 \alpha^2 \sin^2{\theta_{12}}) - {\hat A} w_{cos;2}^{(12)}
\nonumber \\
w_{cos;2}^{(12)} &=& \frac{\alpha}{4} \cos{ \theta_{13} } \sin{2 \theta_{12} } \sin{ 2\theta_{13} }   ( 1 - \alpha \sin^2{ \theta_{12} })  ~,
\eeq

\beq
w_{cos;0}^{(23)} &=&   \frac{\alpha}{8} \cos{ \theta_{13} } \cos{2 \theta_{23} } \sin{2 \theta_{12} } \sin{2 \theta_{13} }    ((1-\alpha)^2 - {\hat A} (  \cos{2\theta_{13} } -  \alpha  (1 - 4 \sin^2{ \theta_{12} } \sin^2{ \theta_{13} }) \nonumber \\       
&+&  2\alpha^2 \sin^2{ \theta_{12} } C^{(-)}_2 )) \nonumber \\
w_{cos;1}^{(23)} &=& \frac{\alpha}{4} \cos{2 \theta_{23} }  \sin{2 \theta_{12} } \sin{ \theta_{13} } 
 ( \cos^2{ \theta_{13} } + \alpha \cos^2{ \theta_{13} }  - 2 \alpha^2 C^{(-)}_2)  \nonumber \\
w_{cos;2}^{(23)} &=& -\frac{\alpha}{2} \cos{2 \theta_{23} } \sin{2 \theta_{12} } \sin{ \theta_{13} }  ( \cos^2{ \theta_{13} }   - \alpha C^{(-)}_2) ~,
\eeq

\beq 
w_{cos;0}^{(22)} &=&  \frac{\alpha}{2} \sin{ \theta_{13} } \sin{2 \theta_{12} } \cos^2{\theta_{13} }   
 ((1-\alpha)^2 \sin^2{ \theta_{23} } - {\hat A} (\cos{2\theta_{13} } \sin^2{ \theta_{23} } \nonumber \\
&+&  \alpha  (  \cos^2{ \theta_{12} } - \sin^2{ \theta_{23} } (1- 4 \sin^2{ \theta_{12} } \sin^2{ \theta_{23} }) )
- 2  \alpha^2 \sin^2{ \theta_{12} } C^{(+)}_1)) \nonumber \\
w_{cos;1}^{(22)} &=&  \frac{\alpha}{2} \sin{2 \theta_{12} } \sin{ \theta_{13} } 
(\sin^2{ \theta_{23} } \cos^2{\theta_{13} }  + \alpha  \cos^2{ \theta_{13} } \sin^2{ \theta_{23} }  - \alpha^2( 1 - 2 C^{(+)}_1) \nonumber \\
&+&  {\hat A}  \cos^2{\theta_{13} } (1 -  \alpha \sin^2 { \theta_{12} }))
\nonumber \\
w_{cos;2}^{(22)} &=& - \frac{\alpha}{2} \sin{2 \theta_{12} } \sin{ \theta_{13} } 
( 2 \cos^2{ \theta_{13} } \sin^2{ \theta_{23} }  -\alpha  ( 1 - 2 C^{(+)}_1 ) ) ~.
\eeq

The coefficients $ w^{(33)}_{cos}    $ are obtained from $ w_{cos}^{(22)}$ by making the replacement $\sin{\theta_{23}} \leftrightarrow \cos{\theta_{23}} $ and by flipping the overall sign. 

\subsection{ Coefficients $ w_{0}^{(mn)} $ }

\beq  
w_{0;0}^{(12)} &=&   \frac{\alpha }{4} (1-\alpha)^2  \sin^2{ \theta_{12} } \sin^2{ 2\theta_{13} } \sin^2{ \theta_{23} }  + \frac{1 }{4} {\hat A} \cos^2{ \theta_{13} } ( \sin^2{2 \theta_{13} } \sin^2{ \theta_{23} } - 4 \alpha \sin^2{ \theta_{12} } \sin^2{ \theta_{13} } \sin^2{ \theta_{23} } \nonumber \\
&\times& ( 2 - 3 \sin^2{\theta_{13} }  ) - 4 \alpha^2 \sin^2{ \theta_{12} }\sin^2{ \theta_{13} } 
( C^{(+)}_1 + \sin^2{ \theta_{23} } \cos{ 2 \theta_{12} } -\sin^2{ \theta_{23} } C^{(-)}_2 +  \sin^2{\theta_{23} } C^{(+)}_2) \nonumber \\
& +&  4 \alpha^3 \sin^2{\theta_{12} } C^{(+)}_1 C^{(+)}_2)
\nonumber \\
w_{0;1}^{(12)} &=& 
-\cos ^2{ \theta_{13}}  (\sin ^2{ \theta_{13}}\sin ^2{ \theta_{23}}
- \alpha \sin ^2{ \theta_{12}}  \sin ^2{ \theta_{13}}\sin^2{ \theta_{23}}
- \alpha ^2 \sin ^2{ \theta_{12}} \sin^2{ \theta_{13}} \sin ^2{ \theta_{23}} \nonumber \\
&+& \alpha^3 \sin ^2{ \theta_{12}} C^{(+)}_1) - {\hat A} w_{0;2}^{(12)}
\nonumber \\
w_{0;2}^{(12)} &=&  \cos ^2{ \theta_{13}}  (\sin ^2{ \theta_{13}}\sin^2{ \theta_{23}}  - 2 \alpha  \sin^2{ \theta_{12}}\sin ^2{ \theta_{13}} \sin ^2{ \theta_{23}}  + \alpha ^2 \sin ^2{ \theta_{12}} C^{(+)}_1)~,
\eeq

The coefficients $ w_{0}^{(13)}$ are obtained from $ w_{0}^{(12)}$ by making the replacement $\sin{\theta_{23}} \leftrightarrow \cos{\theta_{23}} $.  

\beq
w_{0;0}^{(23)} &=& \frac{\alpha}{4} (1 - \alpha )^2 \sin ^2{ 2 \theta_{23}} \cos ^2{ \theta_{13}} C^{(-)}_2 \nonumber \\
&-& \frac{1}{4} {\hat A} \cos ^2{ \theta_{13}}  
(\cos ^2{ \theta_{13}} \sin ^2{ 2 \theta_{23}}\sin ^2{ \theta_{13}}
- \alpha  \sin ^2{ \theta_{13}}  \sin ^2{ 2 \theta_{23}}
(1 + \sin^2{ \theta_{12}} -3 \sin^2{ \theta_{12}} \sin ^2{ \theta_{13}} ) 
\nonumber \\
&-& \alpha^2  (\sin ^2{ 2 \theta_{12}} \sin ^2{ \theta_{13}} + \sin ^2{ \theta_{12}} \sin ^2{ 2 \theta_{23}}   ( 1 - 3\sin^2{ \theta_{13}} ) C^{(-)}_2) \nonumber \\
&+&  \alpha ^3 \sin ^2{ \theta_{12}}  ( \sin ^2{ 2\theta_{12}} \sin ^2{ \theta_{13}} + \sin ^2{ 2 \theta_{23}} + C^{(-)2}_2 ) )
\nonumber \\
w_{0;1}^{(23)} &=& -\frac{1}{4} ( \cos ^4{ \theta_{13}} \sin ^2{ 2 \theta_{23}} 
- \alpha \cos ^2{ \theta_{13}} \sin ^2{ 2 \theta_{23}} C^{(-)}_2
\nonumber \\
&-& \alpha^2 \cos^2{ \theta_{13}} \sin ^2{ 2 \theta_{23}} C^{(-)}_2
+  \alpha^3 ( \sin ^2{ 2 \theta_{12}} \sin ^2{ \theta_{13}} + \sin ^2{ 2 \theta_{23}} C^{(-)2}_2   )  )
\nonumber \\
w_{0;2}^{(23)} &=&    \frac{1}{4} ( \cos^4{\theta_{13}} \sin ^2 {2 \theta_{23} } 
- 2\alpha C^{(-)}_2 \cos^2{\theta_{13}} \sin ^2 {2 \theta_{23} } + \alpha ^2  ( \sin^2 { 2 \theta_{12} } \sin^2 {\theta_{13} } \nonumber \\
&+& \sin^2 {2 \theta_{23}} C^{(-)2}_2) ) ~,
\eeq

\beq 
w_{0;0}^{(11)} &=&  - \frac{\alpha}{4} (1-\alpha )^2 \sin ^2{ \theta_{12}} \sin ^2{ 2 \theta_{13}}  
- \frac{1}{4} {\hat A} \cos ^2{ \theta_{13}}  ( \sin^2{ 2 \theta_{13}}   - 2 \alpha \sin ^2{\theta_{12}}  \sin ^2{ \theta_{13}} ( 1 + 3 \cos{ 2\theta_{13}} )
\nonumber \\
&-& 4 \alpha ^2 \sin^2 {  \theta_{12} } \sin^2 {\theta_{13} }   (1 +  \cos^2{ \theta_{12}} - \sin ^2{ \theta_{12}}  ( 2 - 3 \sin^2{ \theta_{13}} ) ) 
+ 4 \alpha ^3 \sin ^2{ \theta_{12}} C^{(+)2}_2 ) 
\nonumber \\
w_{0;1}^{(11)} &=&  \cos ^2{ \theta_{13}} (  \sin ^2{ \theta_{13}} - \alpha \sin^2{ \theta_{12}}\sin^2{ \theta_{13}} -  \alpha^2 \sin ^2{ \theta_{12}} \sin ^2{ \theta_{13}}  + \alpha^3 \sin ^2{ \theta_{12}} C^{(+)}_2  ) \nonumber \\
&-&   {\hat A} w_{0;2}^{(11)} 
\nonumber \\
w_{0;2}^{(11)} &=& - \cos^2{ \theta_{13}}( \sin^2{ \theta_{13}} -2 \alpha \sin^2{ \theta_{12}} \sin^2{ \theta_{13}} + \alpha ^2  \sin^2{ \theta_{12}} C^{(+)}_2  )
\eeq

\beq 
w_{0;0}^{(22)} &=&   - \alpha (1-\alpha )^2 \cos^2{ \theta_{13}}  \sin^2{ \theta_{23}} C^{(+)}_1 - {\hat A}  \cos ^2{ \theta_{13}}  
 ( \cos^2{ \theta_{13}} \sin^2{ \theta_{13}} \sin^4{ \theta_{23}} \nonumber \\
&-&  \alpha  \sin ^2{ \theta_{13}} \sin ^2{ \theta_{23}}
(1  - 3 C^{(+)}_1 + \cos{ 2 \theta_{12}} \cos{ 2 \theta_{23}} ) \nonumber \\
&+& \alpha ^2 \sin^2{ \theta_{12}} \sin^2{ \theta_{23}} ( 1- 3 \sin^2{ \theta_{13}} ) C^{(+)}_1 + \alpha ^3 \sin ^2{ \theta_{12}} C^{(+)2}_1) \nonumber \\
w_{0;1}^{(22)} &=&  \cos^2{ \theta_{13}} \cos^2{ \theta_{23}} (1 -\cos ^2{ \theta_{13}} \sin ^2{ \theta_{23}})  - \alpha \cos ^2{ \theta_{13}} \sin ^2{ \theta_{23}} C^{(+)}_1 - \alpha^2 \cos ^2{ \theta_{13}} \sin ^2{ \theta_{23}} C^{(+)}_1 + \alpha^3 C^{(+)}_1\nonumber \\
&\times&  ( 1 - C^{(+)}_1)
+   {\hat A}   \cos ^2{ \theta_{13}} ( \sin^2{ \theta_{13}} \sin ^2{ \theta_{23}} -2 \alpha \sin^2{ \theta_{12}} \sin ^2{ \theta_{13}}\sin^2{ \theta_{23}}  +\alpha ^2 \sin^2 \theta_{12} C^{(+)}_1 ) \nonumber \\
w_{0;2}^{(22)} &=& - \cos ^2{ \theta_{13}} \sin ^2{ \theta_{23}} (1 -  \cos ^2{ \theta_{13}} \sin ^2{ \theta_{23}}) + 2 \alpha \sin ^2{ \theta_{23}} (
\cos^2{ \theta_{12}} \cos^2{ \theta_{23}} - \sin^2{ \theta_{23}} \sin^2{ \theta_{12}}  \sin^4{ \theta_{13}} \nonumber \\
&-&  \sin ^2{ \theta_{13}} (\cos ^2{ \theta_{12}}  - \sin ^2{ \theta_{23}})) -  \alpha^2 C^{(+)}_1 ( 1 - C^{(+)}_1)
\eeq

The coefficients $ w^{(33)}_{0}$ are obtained from $ w^{(22)}_{0}$ by making the replacement $\sin{\theta_{23}} \leftrightarrow \cos{\theta_{23}} $.

\end{widetext}

\end{document}